\documentclass[letter]{aa} 
\usepackage{graphicx}
\usepackage{txfonts}
\begin{document}
   \title{The blazar-like radio structure of the TeV source IC\,310}


   \author{Matthias Kadler
          \inst{1}
          \and
          Dorit Eisenacher\inst{1}
	 \and
	Eduardo Ros\inst{2,3}
	\and
Karl Mannheim\inst{1}
\and
Dominik Els\"asser\inst{1}
\and
Uwe Bach\inst{3}
          }

   \institute{Lehrstuhl f\"ur Astronomie, Universit\"at W\"urzburg, 
Campus Hubland Nord, Emil-Fischer-Stra\ss e 31, D-97074 W\"urzburg, Germany\\
              \email{matthias.kadler@astro.uni-wuerzburg.de, dorit.eisenacher@physik.uni-wuerzburg.de,\\
mannheim@astro.uni-wuerzburg.de, elsaesser@astro.uni-wuerzburg.de}
	\and
	 Departament d'Astronomia i Astrof\'{\i}sica, Universitat de Val\`encia, E-46100 Burjassot, Val\`encia, Spain\\
	\email{Eduardo.Ros@uv.es}
         \and
	Max-Planck-Institut f\"ur Radioastronomie, Auf dem H\"ugel 69, D-53121 Bonn, Germany\\
	\email{ubach@mpifr.de}
             }

   \date{Received 6 October 2011 / Accepted 23 December 2011}

 
  \abstract
{The radio galaxy IC\,310 in the Perseus cluster
has recently been detected in the gamma-ray regime at GeV and TeV energies.
The TeV emission shows time variability and an extraordinarily hard spectrum, even harder than the spectrum of the
similar nearby gamma-ray {emitting} radio {galaxy}  M87. 
   }
   {High-resolution studies of the radio morphology help to constrain the geometry of the jet
on sub-pc scales and to find out where the high-energy emission might come from.
   }
   {We analyzed May 2011 VLBA data of IC\,310 at a wavelength of 3.6\,cm, revealing the parsec-scale radio structure of this source.  We compared our 
   findings with more information available from contemporary single-dish flux density measurements 
   with the 100-m Effelsberg radio telescope.
   }
   {We have detected a one-sided core-jet structure with blazar-like, 
   beamed radio emission oriented along the same position angle as the kiloparsec scale radio structure
   observed in the past by connected interferometers.  {Doppler-boosting favoritism} is consistent with an  angle
of $\theta \la 38^\circ$  between
  the jet axis and the {line-of-sight,} i.e., very likely within the boundary dividing low-luminosity radio galaxies and 
BL Lac objects in unified schemes.  }
   { The stability of the jet orientation from parsec to kiloparsec scales in IC\,310 argues against its
   classification as a head-tail radio galaxy; i.e., there is no indication of an interaction with the
intracluster medium that would determine the direction of the tail.  IC\,310 seems to represent
 a low-luminosity FRI radio galaxy at a borderline angle to reveal its   BL Lac-type central engine.}


   \maketitle
%

\section{Introduction}

The S0 galaxy \object{IC\,310} (\object{B0313+411}, \object{J0316+4119}) is located in the Perseus
Cluster at a redshift of $z=0.0189$.  Due to its kiloparsec-scale radio morphology, it has 
been classified as a head-tail or, more specifically, narrow-angle tail radio galaxy (see Sijbring et al.\ \cite{sij98} and Feretti
et al.\ \cite{fer98}).  The radio emission extends up to a distance of projected 400\,kpc. 
Characteristic for the head-tail radio galaxies, which are found only in clusters of galaxies,
is that the jet direction is determined by the galaxy's motion through the intracluster medium.
The ``head'' then marks the bow shock thanks to the  impact of the jet on the  intracluster medium,
and the tail corresponds to the redirected extended jet.
{Along with NGC\,1265, IC\,310 was in fact the first source classified as a tailed radio galaxy (Ryle and Windram \cite{ryl68}) since its narrow, elongated large-scale radio
structure  roughly points radially away from 
the center of the Perseus Cluster.}\\
\vspace*{-0.5\baselineskip}

The MAGIC collaboration (Mariotti \cite{mar10}) 
reported the discovery of very high-energy (VHE) 
gamma-rays at photon energies $E>100$\,GeV  from IC\,310 between
October 2009 and February 2010. Neronov et al.\ (\cite{ner10}) reported
on a \textit{Fermi}/LAT detection above 30\,GeV and the source is included in the 
\textit{Fermi} Large Area Telescope second source catalog (The Fermi-LAT Collaboration\ \cite{2fgl}),
as well as in the second catalog of active galactic nuclei (AGN) detected by the \textit{Fermi} 
Large Area Telescope (The Fermi-LAT Collaboration\ \cite{2lac}).   The origin of the gamma-ray emission
has remained elusive, 
{with the putative bow shock, the jet, or the core considered} as the possible sites.
The VHE spectral slope is extremely hard ($2.00\pm0.14$) and the source
shows gamma-ray variability (Aleksic et al. \cite{ale10}).  \\
\vspace*{-0.75\baselineskip}

IC\,310  is the fourth closest AGN detected so far
at VHE gamma-rays, after Centaurus\,A ($z=0.00183$), \object{M\,87} ($z=0.004$), and
\object{3C\,84} (a.k.a.\ as \object{NGC\,1275}; $z=0.017559$,  
belonging to the Perseus cluster as well,
and also detected by \textit{Fermi}/LAT
).  
M87 in the Virgo Cluster shares many  properties similar to IC\,310, and here
 the emission could be traced to a blazar-like central engine  (Aharonian et al.\ \cite{aha06}, Acciari et al.\ \cite{acc09}).
Unification models of radio-loud active galactic nuclei predict low-luminosity FRI radio galaxies to be the parent
population of 
BL Lacertae objects whose characteristics  would only become dominant for small jet angles to the line-of-sight, although
other effects (e.g., hidden emission line regions, adiabatic losses, mass entrainment, jet deceleration, spine-sheath structure) 
complicate the picture (Xu et al. \cite{xu00}).  



Throughout this Letter, we use the present standard cosmological model
for which 1\,arcsec in the sky 
corresponds to a distance of 0.361\,kpc at the distance of IC\,310 (see Wright \cite{wri06}).


%
  \begin{figure}[t]
   \centering
   \includegraphics[width=\columnwidth]{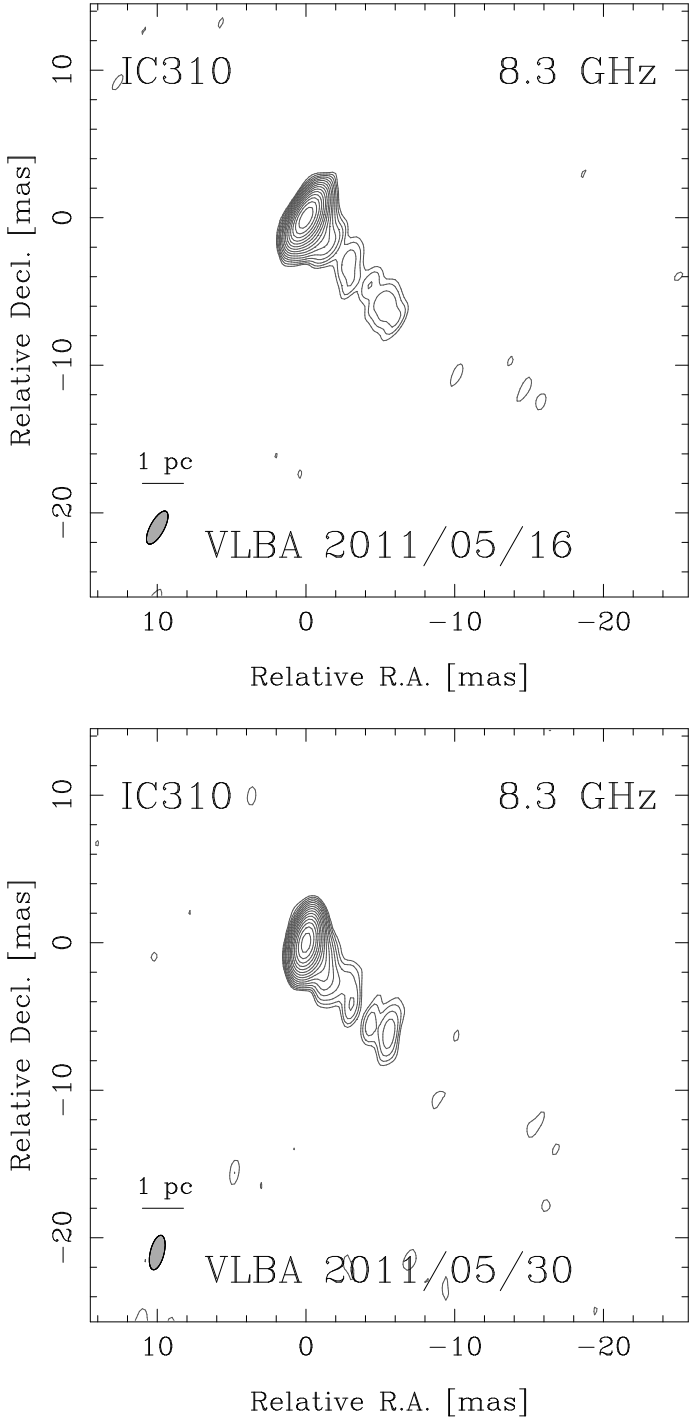}
      \caption{VLBA images from \object{IC\,310} 
taken on May 16 (top) and May 30 (bottom), 2011. 
The image parameters are given in Table~\ref{table:vlba}.
Lowest contours are $2.5 \times$ the rms level.
              }
         \label{fig:VLBAMaps}
   \end{figure}

\section{Observations}
\subsection{Effelsberg} 
{Motivated by the high-energy detection of the putative head-tail radio galaxy IC\,310} (Mariotti \cite{mar10}), we performed a 
run of single-dish test observations with Effelsberg on July 23, 2011
at five frequencies between 2.64\,GHz and 14.60\,GHz (see Table~\ref{table:eff}).
We cross-scaned the source, together with some 
primary and secondary calibrators (namely, \object{3C\,286}, \object{NGC\,7027},
 and \object{3C\,48}).  
{Depending on the wavelength the integration times varied from 20 sec to 60 sec with longer integrations at shorter wavelengths.} 
Our results are presented in Sect.~\ref{subsec:effres}.

\subsection{VLBA}
We have processed archival data (experiment codes 
\texttt{BC196Q} and \texttt{BC196R})
from the Very Long Baseline Array (VLBA)
taken at a wavelength of 3.6\,cm on May 16 and 30, 2011, respectively.  
Those data were correlated at the NRAO Array Operations Center
using the DiFX software corrrelator (see
Deller et al.\ \cite{del11}).  
Our target source was observed during 5\,min at each observing run {for a program to observe 2MASS galaxies (Condon et al. \cite{condon11})}. 
Data were recorded with a {ultra-wide frequency setup, using eight sub-bands of 16\,MHz each spread
over a range from 7.9\,GHz to 8.9\,GHz, with a data bit 
rate of 512\,Mbps.} 
After downloading the data from the archives, we performed the
phase and amplitude calibration using standard methods within
the \textsc{aips} software package.  We
used the measurements of system temperatures and the gain curves
of the stations to calibrate the visibility amplitude
(task \textsc{apcal}), after
doing the digital sampling correction (task \textsc{accor}).
We corrected for the  instrumental phase and delay offsets using
pulse-calibration data recorded during the observations
{at each antenna} (task \textsc{pccor}).
{We obtained the group-delay and phase-rate calibration for each sub-band from a fringe search (task \textsc{fring})}
We got a high detection rate so that no phase-referencing was needed to detect our target source.  The accumulation time used at the correlator was of 0.25\,s, but we averaged the data to 4\,s segments without loss of information or time smearing.

   \begin{table}
      \caption[]{Effelsberg flux density measurements }
         \label{table:eff}
\[
\centering
\resizebox{0.99\columnwidth}{!}{%
\begin{tabular}{@{}lrrr@{\,$\pm$\,}lr@{\,$\pm$\,}lr@{\,$\pm$\,}lr@{\,$\pm$\,}l
@{}}
            \hline
            \hline
            \noalign{\smallskip}
Band & $\lambda$ & \multicolumn{1}{c}{$\nu$} 
& \multicolumn{2}{c}{$I$} 
& \multicolumn{2}{c}{$P$} 
& \multicolumn{2}{c}{$m$} 
& \multicolumn{2}{c}{$\chi$} 
\\
& \multicolumn{1}{c}{[cm]}     & \multicolumn{1}{c}{[GHz]} 
& \multicolumn{2}{c}{[mJy]} 
& \multicolumn{2}{c}{[mJy]} 
& \multicolumn{2}{c}{[\%]} 
& \multicolumn{2}{c}{[deg]} 
\\
            \noalign{\smallskip}
            \hline
            \noalign{\smallskip}
S &  11 &  2.64 & 383 &  3 & 17 & 4 & 4.5 & 1.1 & $-2$ & 7 \\
C &   6 &  4.85 & 238 &  4 & 10 & 4 & 4.0 & 1.7 & $-29$ & 9 \\
X & 3.6 &  8.35 & 148 &  3 & \multicolumn{2}{c}{{$<10$}} & \multicolumn{2}{c}{...} & \multicolumn{2}{c}{...} \\ 
X & 2.8 & 10.45 & 155 & 15 & \multicolumn{2}{c}{{$<10$}} & \multicolumn{2}{c}{...} & \multicolumn{2}{c}{...} \\ 
U &   2 & 14.60 & 103 &  6 & \multicolumn{2}{c}{...} & \multicolumn{2}{c}{...} & \multicolumn{2}{c}{...} \\
            \noalign{\smallskip}
            \hline
   \end{tabular}
}
\]
   \end{table}

   \begin{table}
      \caption[]{VLBA map parameters 
}
         \label{table:vlba}
\[
\centering
\resizebox{0.99\columnwidth}{!}{%
\begin{tabular}{@{}lr@{\,$\pm$\,}lr@{\,$\pm$\,}lcr@{\,$\times$\,}l@{\,,\,}l
@{}}
            \hline
            \hline
            \noalign{\smallskip}
Epoch 
& \multicolumn{2}{c}{$S_\textrm{tot}$} 
& \multicolumn{2}{c}{$S_\textrm{peak}$} 
& \multicolumn{1}{c}{rms} 
& \multicolumn{3}{c}{Beam}
\\
& \multicolumn{2}{c}{[mJy]} 
& \multicolumn{2}{c}{[mJy/beam]} 
& \multicolumn{1}{c}{[mJy/beam]} 
& \multicolumn{3}{r}{[mas\,$\times$\,mas\,,\,deg]} 
\\
            \noalign{\smallskip}
            \hline
            \noalign{\smallskip}
2011/05/16  &  118.8 &  6.0 & 88.7 & 4.4 & 0.20 & 2.50 & 0.94 & $-$28.4 \\
2011/05/30  &  115.0 &  6.1 & 85.1 & 4.6 & 0.21 & 2.39 & 0.94 & $-$13.5 \\
Comb.       &  115.6 &  7.9 & 84.8 & 5.7 & 0.16 & 2.34 & 0.93 & $-$21.5 \\
Comb.\,Taper&  113.6 &  7.7 & 94.5 & 6.4 & 0.17 & 2.63 & 1.76 & $-$21.5 \\
            \noalign{\smallskip}
            \hline
   \end{tabular}
}
\]
   \end{table}

\section{Results}

\subsection{Effelsberg Radio Spectrum\label{subsec:effres}}

From the cross-scan observations, source flux densities were determined following the
procedures described by Kraus et al.\ (\cite{kra03}).
{From 2.64\,GHz to 10.45\,GHz, the measured full width half maximum (FWMH) in elevation was somewhat larger than the nominal antenna beam width, indicating that the source was marginally resolved along the main jet axis (compare Feretti et al. 1998 for an Effelsberg map of IC310 at 10.6GHz). At the highest frequency of 14.6 GHz IC310 appeared compact.} 
The results are shown in Table~\ref{table:eff}, displaying the name of the band, the observing wavelength and frequency, total flux density $I$, linearly polarized flux density $P$, polarization degree $m$, and electric vector position angle $\chi$.

{{\object{IC\,310} exhibited a steep radio spectrum with the flux density $S$, 
which is described well by a power law $S\propto \nu^\alpha$ with a spectral index $\alpha=-0.75$. From previous
studies with the Westerbork and Effelsberg telescopes} (Sijbring \& de Bruyn \cite{sij98}, Feretti et al. \cite{fer98}), it is known that this behavior comes from the optically thin emission of the 
kiloparsec-scale radio jet of \object{IC\,310}, which dominates the spectrum at low frequencies.

At 8.3\,GHz (the frequency of the VLBA observations), we find a total radio flux density of $(148\pm3)$\,mJy. 
A comparison with the total flux density emitted on parsec scales and recovered by  the VLBA
observations (see below) shows that $>80$\,\% of the emission at this high frequency is created
on scales smaller than 1 milliarcsecond.

%
%

  \begin{figure*}[t]
\vbox{
   \includegraphics[width=0.63\textwidth]{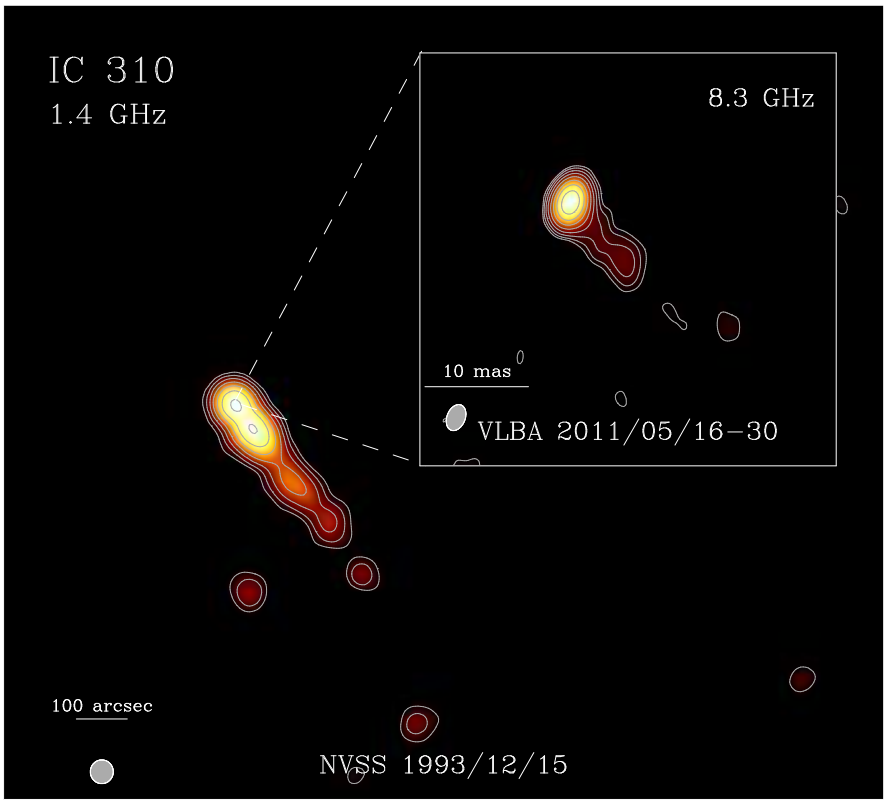}
}\vspace{-5cm}
\hfill \parbox[b]{5.7cm}{
      \caption{The jet in \object{IC\,310} on small and large scales.  Shown are the NVSS image  
(large field; VLA 1.4\,GHz, December 15th, 1993)
and the tapered VLBA image (inset panel; 7.9--8.8\,GHz, combining data 
from May 16 and May 30, 2011).
The convolving beam of the NVSS image is 45$\times$45\,arcsec.
The lowest contour and the brightness peak of the NVSS image are
3.5\,mJy\,beam$^{-1}$ and 173.7\,mJy\,beam$^{-1}$, respectively.
Image parameters for the VLBA image are given in Table~\ref{table:vlba}.
The lowest contour of the VLBA image is $2.5 \times$ the rms value. 
              }
         \label{fig:collage}
}
   \end{figure*}

\subsection{VLBA imaging results}
After \textsc{aips} processing, we exported our data into the
software package \textsc{difmap}, where we performed hybrid mapping
using the CLEAN algorithm.  Since the a priori amplitude
calibration was excellent, as seen in other calibrators used
during the observations (i.e., NGC\,1052, 0059+581, 2200+420), we
started our processing by only phase self-calibration.
During the imaging process, we tested and applied different weighting schemes 
(natural weighting, uniform weighting, and natural weighting
with downweighting the longest baselines, tapering).
Because of the sparse $(u,v)$-coverage, we restricted the CLEAN-region
to an area of 8 to 10 beam sizes to keep the number of CLEAN
components small, hence ensuring that the number of degrees of freedom
can be supported by the data. 
In all cases
we obtained a core-jet morphology with extended emission 
oriented towards the southeast with a position angle of 
$-135^\circ$, reaching up to 20\,mas and with hints of emission
up to 40\,mas away from the brightest component.
The images resulting from this process, with natural weighting
and full resolution, are presented in Fig.~\ref{fig:VLBAMaps}.

We continued our analysis by applying gentle amplitude self-calibration (typical corrections are smaller than 10\%).  
This,
together with the short time observed (5.3\,min each time) makes
the quality of the images  quite remarkable, with a dynamic range of $\sim150:1$.

Both data sets on May 16 and May 30 had similar flux density measured 
at the shortest VLBA baseline between Los Alamos and Pie Town.
Comparing the two independent images, we find that no significant variability occurred between
both epochs, so we merged both 5-min data sets (\textsc{aips} task 
\textsc{dbcon}) to get an image with a 
higher signal-to-noise ratio.  In Figure~\ref{fig:collage}, we show the results
of hybrid mapping with a Gaussian 
tapering at 125\,M$\lambda$, 
together with an image of the
NRAO VLA Sky Survey (NVSS; Condon et al.\ \cite{con98})
taken on December 1993 with a resolution of 45\,arcsec.  
The resulting VLBA beam size projected along the jet direction is $\sim2$\,mas,
whereas the  resolution obtained with natural weighting and considering
all baselines was $\sim1$\,mas in size.
The expected thermal noise for the full VLBA at
512 Mbps for 5.3\,(10.6)\,min is of 0.229\,(0.162)\,mJy/beam (1-$\sigma$ level).
Our formal image root-mean-square (rms) noise level reaches these
theoretical values (see Table~\ref{table:vlba}) but does not exceed them significantly, verifying that no \textit{overcleaning}
occurred. 

Gaussian model fitting the interferometric visibilities of all three data sets
yields a core flux density emission of $80-90$\,mJy, and
a fitted size of (0.3--0.4)\,mas, which corresponds  
to a brightness temperature on the order
of 1.5$\times$10$^{10}$\,K.  The jet components
are responsible for additional 14\,mJy at a distance of 0.8\,mas in
P.A.\ $-144^\circ$, and then an additional 10\,mJy in the range
of 3\,mas to 8\,mas away from the core; i.e., the compactness
of the source from the 8.3\,GHz snapshot images 
is $\sim 70$\,\%.

%

\section{Discussion}

Figure~\ref{fig:collage} shows that the parsec-scale jet obtained with the
VLBA snapshot images and the kiloparsec-scale jet  from the
NVSS  are oriented
in the same direction.  That is, there is no misalignment
over several orders of magnitude in the scale of the jet. In more
powerful and more distant sources, {the} one-sidedness of jets
is usually attributed to relativistic bulk motion along a relatively small
angle to the line of sight, which leads to
Doppler boosting of the jet and deboosting of the 
counterjet emission. With no evidence of any counterjet emission on any scale,
we can assume that the peak signal at 8.3\,GHz recovered by the VLBA
may be attributed to the jet and that the counterjet signal is suppressed to a
value below $3 \sigma$, with $\sigma$ the root-mean-square noise in the
final VLBA image. Under these assumptions, the jet-to-counterjet ratio is $R>177$. 
Assuming now that Doppler-boosting is responsible for the one-sidedness,
{which is generally considered to be} the main factor on the pc-scale in unified models for radio galaxies and BL Lacs,
$R$ is related to the jet speed $\beta$ and the angle to the
line of sight $\theta$ by
\begin{equation}
R=\left(\frac{1+\beta \cos{\theta}}{1-\beta \cos{\theta}}\right)^{2-\alpha}=(\beta_\mathrm{app}^2+\delta^2)^{2-\alpha} \quad ,\label{eq1}
\end{equation}
with the spectral index $\alpha$ defined through $S\propto \nu^\alpha$, {the apparent jet speed $\beta_\mathrm{app}=v_\mathrm{app}/c$, and the Doppler factor $\delta$ (Urry \& Padovani \cite{sij98})}.  Further studies will {measure $\beta_\mathrm{app}$ and} elucidate if free-free absorption or adiabatic losses can also play a role in this source.
In the case of Doppler-boosting favoritism, the measured limit on the jet-to-counterjet ratio leads to an upper limit
on $\theta$:
\begin{equation}
\theta < \arccos{\left(\frac{R^{\frac{1}{2-\alpha}}-1}{R^{\frac{1}{2-\alpha}}+1}\right)} \quad \beta^{-1} \quad .\label{eq2}
\end{equation}

Attributing all of the 8.3\,GHz VLBA peak flux density to the jet, assuming $\alpha=0$ (for flat-spectrum compact emission), and letting $\beta \rightarrow 1$ lead to
$\theta < 31^\circ$. A somewhat more conservative estimate is derived from using the emission maximum
3\,arcmin down the jet seen in the 49\,cm WSRT image of Sijbring et al.\ (\cite{sij98}). Knowing now that
IC\,310 shows a one-sided jet morphology on all scales, this local peak of $\sim400$\,mJy compared to the
$3 \sigma$ noise on the opposite side of the core of $\sim0.8$\,mJy leads to a jet-to-counterjet ratio
of $R>500$. Using the spectral index $\alpha=-0.75$ measured at this position, the angle to the line of sight is constrained to $\theta<36^\circ$. A third way to constrain the jet angle to the line of sight is to consider 
the first jet maximum in the NVSS image at 1.4\,GHz located at about 75$^{\prime\prime}$ ($\sim 23$\,kpc) 
downstream from
the core of the jet (see Fig.~\ref{fig:collage}). Here, we derive a jet-to-counterjet ratio of $162/0.8=202$.
With a spectral index of $-0.5$ at this position, the angle to the line of sight is constrained to
$\theta < 38^\circ$. {For these three scenarios ($R>177,\alpha=0$; $R>500,\alpha=-0.75$; $R>202,\alpha=-0.5$) and an assumed apparent jet speed of $\beta_\mathrm{app}=1$, which would be typical of low-luminosity high-peaked BL\,Lac objects (Piner \& Edwards \cite{piner04}, Piner et al. \cite{piner10}), Eq.~(\ref{eq1}) yields minimum Doppler factors of $\delta>3.5$, $\delta>2.9$, and $\delta>2.7$, respectively. Thus, the Doppler factor of IC\,310 may be well above the typical values estimated for Cen\,A (e.g., Abdo et al. \cite{abdo10}: $\delta<3.8$), M\,87 (e.g., Abdo et al. \cite{abdo09a}: $\delta\sim3.9$), or NGC\,1275 (e.g., Abdo et al. \cite{abdo09b}: $\delta\sim2.3$). Based on the available data, the}  jet morphology and the jet-to-counterjet ratio point towards a blazar-like nature 
for IC\,310, rather than radio-galaxy characteristics, {although higher-signal-to-noise-ratio VLBI observations are needed to put stronger constraints on both $R$ and $\delta$}.

The blazar-like nature of IC\,310 is supported by the findings of Owen, Ledlow \& Keel (\cite{owe96}) and Rector, Stocke \& Perlman (\cite{rec99}). In their
optical spectroscopic survey of 190 objects {in the Abell clusters}, the former find four sources with weak nonthermal
activity, which they identify as BL\,Lac candidates. IC\,310 is one of these four objects. The difficulty of
detecting the optical BL\,Lac in nearby low-luminosity objects like IC\,310 is due to the dominance of the
host-galaxy star light. At X-ray energies, Sato et al. (\cite{sat05}) find a featureless power-law spectrum
from an unresolved point source at the position of the IC\,310 nucleus (embedded in strong extended thermal
cluster emission) with a rather steep photon index of $\Gamma=2.5$ and suggest a possible BL\,Lac type
nature of this emission as well, although no variability at X-rays has so far been reported.
The two-point spectral index between the VHE flux and the (noncontemporaneous)  X-ray flux 
$\alpha_{\rm X/VHE}\sim 1$ is
also in line with an underlying blazar-type spectral energy distribution (Albert et al. \cite{alb08}).

\section{Conclusions}

      We analyzed archival snapshot 8.3\,GHz observations from the VLBA taken in May 2011, and found a core-jet subparsec scale structure in the emission of \object{IC\,310}.  The source is core-dominated, with a flux density of around 0.1\,Jy, and there is extended emission towards the SE, in P.A.\ $-135^\circ$, at the same direction of the kiloparsec-scale emission as observed by the VLA in the NVSS survey (Condon et al.\ \cite{con98}) or the WSRT (Feretti et al.\ \cite{fer98}).
      
Single-dish continuum spectra from the 100-m telescope in 
Effelsberg taken in late July 2011 reveal radio emission at the level of
0.1\,Jy at 15\,GHz to 0.4\,Jy at 2.6\,GHz.  The value of the flux density at the VLBA observing frequency is  $\sim 0.15$\,Jy, which implies that the extended emission not detected with the VLBA is smaller than 25\%, provided that no
significant radio variability at this frequency occurred between May and July
2011.

There is no evidence of interactions of the kpc jet with the intercluster medium as responsible for
its  narrow-tail morphology; instead, we found  evidence that its orientation is inherited from the
pc-scale inner jet. {It therefore follows that IC\,310} should not be classified as a head-tail radio galaxy.

The orientation angle of the jet is consistent with a low-luminosity radio galaxy at the borderline to  BL Lacs
assuming relativistic bulk motion. This finding corroborates evidence that the high-energy emission of IC\,310
originates in the central blazar-like engine.

\begin{acknowledgements}
We thank Stefan R\"ugamer and A.\,P.\,Lobanov for valuable discussions and critical reading of the manuscript. {We also thank the referee for constructive suggestions.}
The Very Large Array and the Very Long Baseline Array are operated
by the National Radio Astronomy Observatory, a facility of the National
Science Foundation operated under cooperative agreement by Associated
Universities, Inc.
This work made use of the Swinburne University of Technology software correlator, developed as part of the Australian Major National Research Facilities Programme and operated under licence. 
The results presented here are based on observations with the 
100-m telescope of the MPIfR (Max-Planck-Institut f\"ur 
Radioastronomie) at Effelsberg.  
This research made use of the NASA/IPAC Extragalactic Database (NED), which is operated by the Jet Propulsion Laboratory, California Institute of Technology, under contract with the National Aeronautics and Space Administration. 
This research made use of NASA's Astrophysics Data System.
D.E. acknowledges support by the German BMBF Verbundforschung, and
E.R. acknowledges
partial support by the Spanish MICINN through grant 
AYA2009-13036-C02-02, and by the COST action MP0905 
``Black Holes in a Violent Universe".
\end{acknowledgements}

\end{document}